# How many scientific papers are mentioned in policy-related documents?

# An empirical investigation using Web of Science and Altmetric data

Robin Haunschild[*] and Lutz Bornmann[**]

[*]First author and corresponding author:

Max Planck Institute for Solid State Research

Heisenbergstr. 1,

70569 Stuttgart, Germany.

Email: R.Haunschild@fkf.mpg.de

[**] Contributing author:

Division for Science and Innovation Studies

Administrative Headquarters of the Max Planck Society

Hofgartenstr. 8,

80539 Munich, Germany.

Email: bornmann@gv.mpg.de

**Abstract**

In this short communication, we provide an overview of a relatively newly provided source of altmetrics data which could possibly be used for societal impact measurements in scientometrics. Recently, Altmetric – a start-up providing publication level metrics – started to make data for publications available which have been mentioned in policy-related documents. Using data from Altmetric, we study how many papers indexed in the Web of Science (WoS) are mentioned in policy-related documents. We find that less than 0.5% of the papers published in different subject categories are mentioned at least once in policy-related documents. Based on our results, we recommend that the analysis of (WoS) publications with at least one policy-related mention is repeated regularly (annually). Mentions in policy-related documents should not be used for impact measurement until new policy-related sites are tracked.

## 1. Introduction

In recent years, bibliometrics (citation analysis) has become the gold standard in measuring impact of the sciences (Committee for Scientific and Technology Policy, 2014). However, it is an important drawback of citation analysis that it only measures the influence of scientific work on science itself. Today, research funders and science politicians are interested in the broad impact of science, i.e. the impact of the sciences beyond the sciences. Thus, scientometricians are searching for new ways of measuring impact. Altmetrics (alternative metrics) might offer these new ways: "Altmetrics are usually based on activity on social media platforms, which relates to scholars or scholarly content … However, altmetrics also comprise mentions in mainstream media or policy-related documents, as well as usage metrics such as full text views and downloads, although these have been available long before the concept of altmetrics was introduced. The common denominator of these heterogeneous metrics is that they exclude, and are opposed to, 'traditional' bibliometric indicators" (Work, Haustein, Bowman, & Larivière, 2015). Thus, altmetrics are intended to complement traditional (citation-based) metrics.

According to Thelwall and Kousha (2015) "in theory, alternative metrics may be helpful when evaluators, funders or even national research assessment need to know about 'social, economic and cultural benefits and impacts beyond academia'" (p. 588; see also Moed & Halevi, 2015). In altmetrics studies, blogging, social bookmarks, and microblogging are of particular interest (Bornmann, 2015). Since a short time ago, many publishers have added altmetrics to the papers published in their journals (Wilsdon et al., 2015). Currently, three bigger companies offer altmetrics data on the paper level: Altmetric (www.altmetric.com), Plum Analytics (www.plumanalytics.com), and Impact Story (www.impactstory.org) (Zahedi, Costas, & Wouters, 2014). Recently, Altmetric has begun to cover another source of altmetrics data which is of particular interest for measuring the broad impact of the sciences: mentions of scholarly papers in policy-related documents: "As you might have already learned from our June press release announcing the launch of Altmetric for Institutions, we recently started tracking some highly impactful new sources of attention: policy and guidance documents" (Liu, 2014).

The definition of societal impact given by Wilsdon, et al. (2015) shows the potential use of mentions of scholarly papers in policy-related documents: "Research has a societal impact when auditable or recorded influence is achieved upon non-academic organisation(s) or actor(s) in a sector outside the university sector itself – for instance, by being used by one or more business corporations, government bodies, civil society organisations, media or specialist/professional media organisations or in public debate. As is the case with academic impacts, societal impacts need to be demonstrated rather than assumed. Evidence of external impacts can take the form of references to, citations of or discussion of a person, their work or research results" (p. 6). Non-academic organizations or actors are often authors of policy documents. Therefore, the newly provided source by Altmetric (mentions of scholarly papers in policy-related documents) might be valuable for the assessment of societal impact.

In this short communication, we test the scope and informative value of the newly provided altmetrics. As a first empirical analysis on a large publication set, we study how mentions of policy-related documents are distributed of scientific fields. We also analyze how many papers published since 2000 have received at least one policy-related mention.

## 2. Data Set

**2.1. Sources for measuring impact in policy-related documents**

There is no clear definition of what a policy-related document is and what is not. For the current study, we adopt the provided list of policy-related document sources by Altmetric. According to Liu (2014) Altmetric tracked over 40 policy-related sites by the end of 2014.

According to the data which we received from Altmetric on December 19, 2015, more than 100 policy-related sources are currently tracked by Altmetric. Some examples for very frequent sources are as follows:

- European Food Safety Authority
- The Association of the Scientific Medical Societies in Germany
- World Health Organization
- Food and Agriculture Organization of the United Nations
- World Bank
- UK Government (GOV.UK)
- National Institute for Health and Care Excellence
- Australian Policy Online
- NIHR Journals Library
- International Monetary Fund
- Intergovernmental Panel on Climate Change

A regional bias can be seen in the current selection of policy-related sites: Mainly, international, English documents are tracked by Altmetric. Non-English policy sources are not covered, yet, but Altmetric continues to broaden their coverage (Konkiel, 2016).

## 2.2. Analyses of policy-related documents

It is common practice in scientometrics to evaluate the impact of articles and reviews. Other document types are usually not included in evaluative bibliometrics (Moed, 2005). We merged the data received from Altmetric with our bibliometric in-house database (which is based on the Web of Science, WoS, data) to analyze the number and percentage of papers (articles and reviews) with a policy-related mention in different WoS subject categories. The combination of our bibliometric database with the data received from Altmetric was only possible via the DOI so that only papers with a DOI are considered. Since 2006, at least half of the papers indexed in the WoS have a DOI, while 83.5% of the altmetrics records have one. We restrict our analysis to the publication years between 2000 and 2014. The results of (Bornmann, Haunschild, & Marx, 2016) have shown that papers published before 2000 are only occasionally mentioned in policy-related documents.

In total 11,254,636 papers from our bibliometric database were used in this study and 35,504 papers (0.32%) with at least one policy-related mention were found. In total, we found papers with at least one policy-related mention in 228 WoS subject categories.

## 3. Results

Table **1** shows the number of papers (with DOI) published per year with the number and percentage of papers with at least one policy-related mention. The data show that policy-related mentions are rather rare events: Less than 0.5% of the papers are cited in policy-related documents at least once. The number and the percentage of papers with a policy-related mention exhibit with 0.48% a maximum in the publication year 2005. This indicates a longer time frame for papers to be mentioned in a policy-related document than for papers to be cited in another scientific paper. The time-curve of the citations referenced by scientific papers usually shows a distinct peak between two and four years after publication of the cited paper (Redner, 2005).

Table 1: Annual number and percentage of papers with at least one policy-related mention

| publication year | #papers | #papers with policy-related mention | %papers with policy-related mention |
|---|---|---|---|
| 2000 | 15,783 | 34 | 0.22% |
| 2001 | 39,553 | 89 | 0.23% |
| 2002 | 226,124 | 645 | 0.29% |
| 2003 | 432,944 | 1,436 | 0.33% |
| 2004 | 541,371 | 2,250 | 0.42% |
| 2005 | 619,459 | 2,943 | 0.48% |
| 2006 | 714,488 | 3,316 | 0.46% |
| 2007 | 789,690 | 3,493 | 0.44% |
| 2008 | 873,244 | 3,702 | 0.42% |
| 2009 | 963,185 | 3,807 | 0.40% |
| 2010 | 1,028,108 | 3,530 | 0.34% |
| 2011 | 1,123,283 | 3,358 | 0.30% |
| 2012 | 1,213,228 | 3,189 | 0.26% |
| 2013 | 1,313,849 | 2,433 | 0.19% |
| 2014 | 1,360,327 | 1,279 | 0.09% |

Table **2** and Table **3** show the number of papers broken down by WoS subject categories (journal sets as defined by Thomson Reuters) together with the number and percentage of papers with at least one policy-related mention. Both tables show the top 20 WoS subject categories with the largest absolute and relative number of papers with policy-related mentions. Table **2** is ordered by the number of papers with at least one policy-related mention while Table **3** is ordered by the percentage of papers with at least one policy-related mention.

Table 2: Number and percentage of papers with at least one policy-related mention broken down by WoS subject categories (decreasingly ordered by the number of papers with at least one policy-related mention). Only the top 20 WoS subject categories are shown.

| WoS subject category | #papers | #papers with policy-related mention | %papers with policy-related mention |
|---|---|---|---|
| Public, Environmental & Occupational Health | 172,404 | 2,730 | 1.58% |
| Economics | 123,866 | 2,695 | 2.18% |
| Medicine, General & Internal | 110,512 | 2,163 | 1.96% |
| Environmental Sciences | 281,019 | 1,873 | 0.67% |
| Microbiology | 183,390 | 1,615 | 0.88% |
| Oncology | 283,949 | 1,582 | 0.56% |

| WoS subject category | #papers | #papers with policy-related mention | %papers with policy-related mention |
| --- | --- | --- | --- |
| Infectious Diseases | 91,525 | 1,404 | 1.53% |
| Toxicology | 96,665 | 1,230 | 1.27% |
| Multidisciplinary Sciences | 236,599 | 1,193 | 0.50% |
| Nutrition & Dietetics | 76,966 | 1,190 | 1.55% |
| Immunology | 193,486 | 1,129 | 0.58% |
| Pharmacology & Pharmacy | 285,497 | 1,107 | 0.39% |
| Gastroenterology & Hepatology | 102,243 | 1,103 | 1.08% |
| Pediatrics | 128,259 | 949 | 0.74% |
| Urology & Nephrology | 101,963 | 945 | 0.93% |
| Veterinary Sciences | 79,379 | 927 | 1.17% |
| Food Science & Technology | 147,440 | 901 | 0.61% |
| Surgery | 300,303 | 860 | 0.29% |
| Parasitiology | 40,899 | 852 | 2.08% |
| Psychiatry | 134,098 | 832 | 0.62% |

Table 3: Number and percentage of papers with at least one policy-related mention broken down by WoS subject categories (decreasingly ordered by the percentage of papers with at least one policy-related mention). Only the top 20 WoS subject categories are shown.

| WoS subject category | #papers | #papers with policy-related mention | %papers with policy-related mention |
| --- | --- | --- | --- |
| Agricultural Economics & Policy | 4,247 | 126 | 2.97% |
| Tropical Medicine | 18,569 | 491 | 2.64% |
| Economics | 123,866 | 2,695 | 2.18% |
| Business, Finance | 25,874 | 555 | 2.15% |
| Parasitiology | 40,899 | 852 | 2.08% |
| Medicine, General & Internal | 110,512 | 2,163 | 1.96% |
| Planning & Development | 20,819 | 382 | 1.83% |
| Health Policy & Services | 31,920 | 514 | 1.61% |
| Public, Environmental & Occupational Health | 172,404 | 2,730 | 1.58% |
| Nutrition & Dietetics | 76,966 | 1,190 | 1.55% |
| Infectious Diseases | 91,525 | 1,404 | 1.53% |
| Substance Abuse | 27,351 | 402 | 1.47% |
| Environmental Studies | 45,490 | 623 | 1.37% |
| Health Care Sciences & Services | 58,420 | 776 | 1.33% |
| Toxicology | 96,665 | 1,230 | 1.27% |
| Allergy | 19,283 | 236 | 1.22% |
| Veterinary Sciences | 79,379 | 927 | 1.17% |
| Primary Health Care | 4,990 | 55 | 1.10% |
| Gastroenterology & Hepatology | 102,243 | 1,103 | 1.08% |
| Social Sciences, Biomedical | 22,030 | 229 | 1.04% |

In agreement with the results in Table **1**, the number and percentage of papers with at least one policy-related mention are generally low in every WoS subject category. As Table **3** reveals, only less than 3% of the papers in the WoS subject categories are cited at least once. The WoS subject categories with the largest absolute and relative number of papers with policy-related mentions in Table **2** and Table **3** are closely related to either medicine or economics. Taken as a whole, all subject categories in the tables have a significant connection to the practical use of scientific results (e.g. Primary Health Care, Allergy, or Business, Finance). These are disciplines where a more significant societal impact is understandable. However, the observed disciplinary differences might also be related to a bias in the selection of policy-related sources by Altmetric. Altmetric tracks the most easily searchable policy-related sources. Policy-related sources which are not that easy to track (e.g., International Atomic Energy Agency or International Organization of Standardization) might be more common in disciplines other than medicine or economics.

## 4. Discussion

We expected to find many papers mentioned in policy documents as to be anticipated by the claim of Khazragui and Hudson (2015) that "it is rare that a single piece of research has a decisive influence on policy. Rather policy tends to be based upon a large body of work constituting 'the commons'" (p. 55). The results of this study reveal that only a small part of papers covered in the WoS has been mentioned in policy-related documents. The percentage of mentioned papers is much higher for Mendeley reader counts (Haunschild & Bornmann, 2016) and also higher for tweets (Bornmann & Haunschild, 2016).

Possible reasons for the low percentage of papers mentioned in policy-related documents are: (1) Altmetric quite recently started to analyze policy documents and the coverage of the literature is still low (but will be extended). (2) Maybe only a small part of the literature is really policy relevant and most of the papers are only relevant for scientists. (3) Authors of policy-related documents often are not researchers themselves. Therefore, a scientific citation style should not be expected in policy-related documents in general. Thus, policy-related documents may not mention every important paper on which a policy-related document is based on. (4) There are possible barriers and low interaction levels between researchers and policy makers. However, our results are not too surprising; they reflect the current state of impact measurements using altmetrics: "Alternative metrics may in the future provide useful insights, but currently they are at a very early stage" (Martin, Nightingale, & Rafols, 2014, p. 5). Wilsdon, et al. (2015) outline that "the systematic use of alternative indicators as pure indicators of academic quality seems unlikely at the current time, though they have the potential to provide an alternative perspective on research dissemination, reach and 'impact' in its broadest sense" (p. 45). Similar assessments of altmetrics can be found in Weller (2015).

Mentions in policy-related documents should not be used for impact measurement until more policy-related sites are tracked by Altmetric (or another data provider). However, once more sources of policy-related documents are tracked normalized indicators might be a reliable way for impact assessments of research on policy as one part of society. For Twitter data which is concerned by a similar problem (on too few papers are tweeted), Bornmann and Haunschild (2016) proposed the use of the 80/20 scientometric data quality rule (Strotmann & Zhao, 2015):

A reliable field-specific study is only possible with a database, if 80% of the field-specific publications are covered in this database.

The main limitations of our study are as follows: (1) We restricted the dataset to articles and reviews with a DOI and (2) used only policy-related sites tracked by Altmetric. (3) It is unknown where a particular publication was mentioned, especially on policy-related websites. The sources can be huge and also contain CVs. Therefore, some policy-related mentions might originate from CVs (i.e. the publications listed in a CV). However, considering the small percentage of WoS publications mentioned in policy-related documents (which we found in this study), we expect that only very few mentions originate from such unintended sites.

Despite the current limitations of the newly provided altmetrics source, we think that policy-related mentions of publications will offer interesting impact analyses in future studies. For example, when more policy-related sites are tracked (by Altmetric or other data providers) one could focus on certain policy-related sources (e.g., only British or health-related sources). Depending on the focus of the impact study, one could study the specific impact of publications on certain nations or political areas (e.g. health policy).

## Acknowledgements

The bibliometric data used in this paper are from an in-house database developed and maintained by the Max Planck Digital Library (MPDL, Munich) and derived from the Science Citation Index Expanded (SCI-E), Social Sciences Citation Index (SSCI), and Arts and Humanities Citation Index (AHCI) prepared by Thomson Reuters (Philadelphia, Pennsylvania, USA). The altmetrics data about policy-related mentions were taken from a data set retrieved from Altmetric on December 19, 2015.